%% file: main.tex
\documentclass[review]{elsarticle}
\usepackage{amsmath}
\usepackage{url}
\usepackage{graphicx}
\usepackage[english]{babel}
\usepackage{subfig}
\usepackage{wrapfig}
\usepackage{multicol}
\usepackage{algorithm2e}

\usepackage[dvipsnames]{xcolor}
\definecolor{p3}{cmyk}{0, 0.7808, 0.4429, 0.1412}
\usepackage[colorinlistoftodos]{todonotes}
\usepackage[colorlinks = true,
            linkcolor = Maroon,
            urlcolor  = NavyBlue,
            citecolor = Maroon,
            anchorcolor = NavyBlue,
            bookmarks=false]{hyperref}

\journal{Future Generation Computer Systems}

\begin{document}
\begin{frontmatter}

\title{Adaptive Edge Process Migration for IoT in Heterogeneous Cloud-Fog-Edge  Computing Environment}


\author[ut1]{Chii Chang\corref{mycorrespondingauthor}}
\cortext[mycorrespondingauthor]{Corresponding author}
\ead{chii.chang@ut.ee}

\author[ut2]{Amnir Hadachi}
\ead{amnir.hadachi@ut.ee}

\author[ut1]{Satish Narayana Srirama}
\ead{satish.srirama@ut.ee}

\address[ut1]{Mobile \& Cloud Lab, Institute of Computer Science, University of Tartu, Estonia}
\address[ut2]{ITS Lab, Institute of Computer Science, University of Tartu, Estonia}

\date{October 2018}

\begin{abstract}

\input{sections/abstract}
\end{abstract}


\end{frontmatter}

\input{sections/introduction}
\input{sections/relatedWorks}

\input{sections/system}
\input{sections/rem_scheme}
\input{sections/evaluation}
\input{sections/conclusion}

\section*{Acknowledgement}
The work is supported by the Estonian Centre of Excellence in IT (EXCITE), funded by the European Regional Development Fund.

\bibliographystyle{elsarticle-num}
\bibliography{fog}
\end{document}

%% file: sections/abstract.tex
The latency issue of the cloud-centric IoT management system has motivated Fog and Edge Computing (FEC) architecture, which distributes the tasks from the cloud to the edge resources such as routers, switches or the IoT devices themselves. Specifically, mobile sensors of IoT system can also carry certain tasks for FEC. Considering the need of dynamic process migration from the mobile sensors to other resources when the mobile sensors unable to continue their tasks, the IoT system needs to provide a flexible mechanism that allows the mobile sensors dynamically migrate their tasks to the other FEC resource at runtime. However, it raises a question in what is the optimal approach when the mobile sensors intend to migrate their tasks to multiple heterogeneous FEC resource? In order to address the question, the authors propose REM scheme, which is capable of optimising the process migration decision. Further, in order to realise such a system and to validate the REM scheme, the authors have developed EPIoT host framework. Finally, the authors have implemented and have evaluated the REM scheme and framework, the results have shown that the REM scheme is capable of enhancing the performance of the process migration in heterogeneous FEC environment.

%% file: sections/introduction.tex
\section{Introduction}
The cloud computing paradigm has motivated various information systems moving from the local to the global Internet-based servers. Hence, having the management systems reside remotely have become a common practice for information systems. Unexceptionally, common Internet of Things (IoT) management systems have also followed the same architecture design \cite{Chang2016:CSUR}. As the IoT front-ends continue increasing, and the involved data is getting larger, the drawback of the distant centralised system has risen, which is known as the latency issues, and it mainly derives from the fact that IoT system relies on the Internet as the main communication protocol between the front-end IoT devices and the back-end management server. Researchers discovered that there is a need to keep a certain process and decision making within or near the network where the front-ends are located in order to enhance the agility of the decision making in IoT applications. Such a paradigm is known as fog computing architecture.

A fog computing server is capable of providing storage, compute, acceleration, networking and control services by utilising virtualisation or containerisation technologies \cite{Chang2018:Intro}. For example, industrial integrated routers, which are capable of providing Virtual Machine (VM) or containers engine mechanisms, can support similar software deployment platform as the common cloud services. Moreover, by extending the Over-The-Air (OTA) programming mechanisms \cite{Rossi2010:OTA}, the IoT system can also distribute certain tasks to the resource constraint devices towards enabling self-managed IoT devices. 

Today, fog computing architecture has become one of the main elements of IoT. Specifically, the market research has specified that the main applications of fog computing will be electricity/utilities, transportation, healthcare, industrial activities, agriculture, distributed datacenters, wearables, smart buildings, smart cities, retail and smart homes \cite{FourFiveOneResearch}. Further, the emerged industrial standard IEEE 1934\footnote{see \url{https://standards.ieee.org/news/2018/ieee1934-standard-fog-computing.html}} and the market research report \cite{FourFiveOneResearch} indicate that the population of fog services will increase and it is foreseeable that fog services will not be limited in private networks but also available as public cloud today, in which different service providers can provide their fog computing servers for general public \cite{Chang2017:IndieFog}, which we can consider it as the public fog.

\begin{figure}
  \centering
    \includegraphics[width=0.80\textwidth]{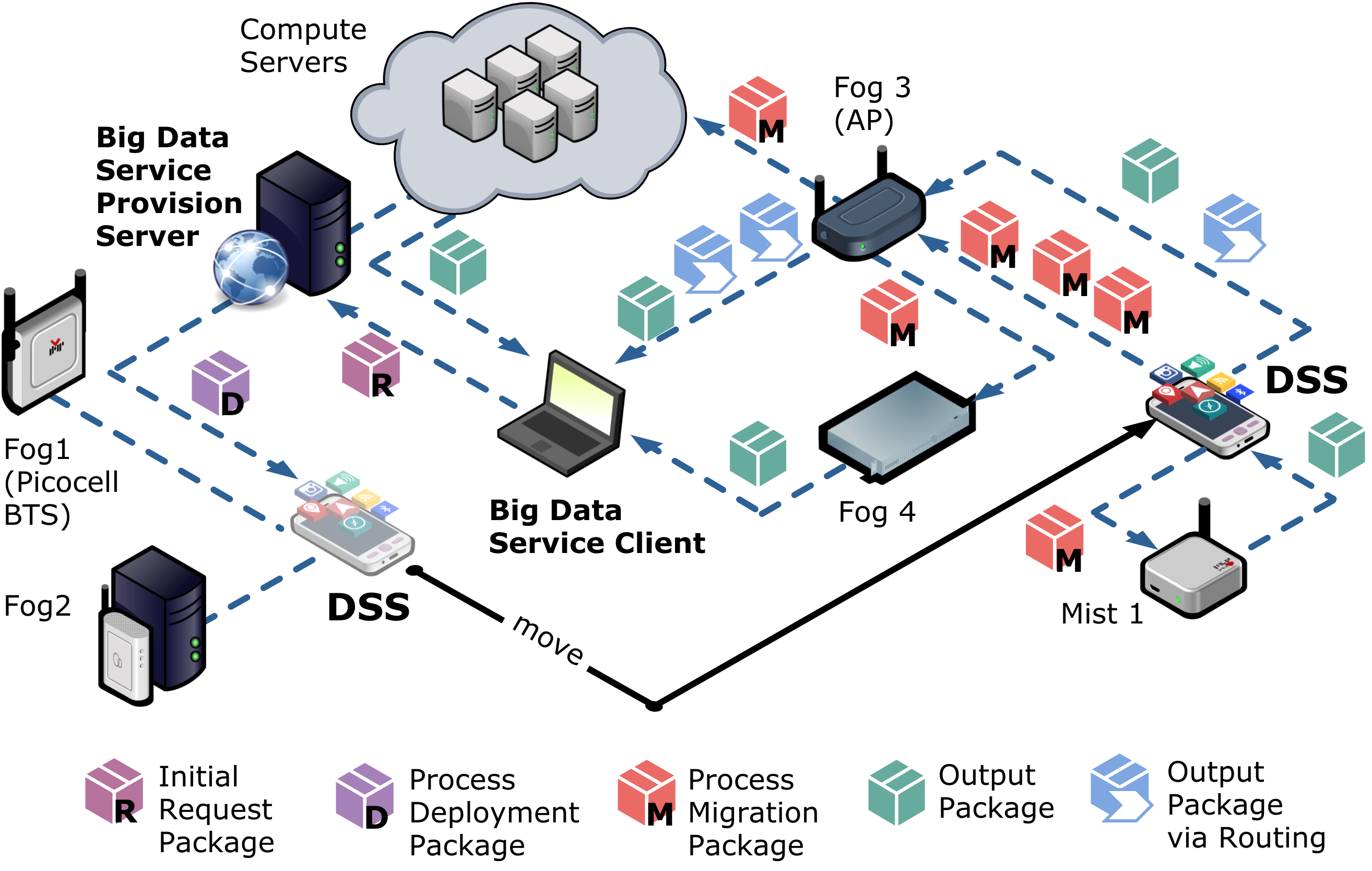}
    \caption{Distributed processing in heterogeneous fog computing environment.}
    \label{fig:scen}
\end{figure}

Suppose the near future smart cities encompass many local service providers who provide various public fog features, performing fog computing is no longer requiring IoT system management team to deploy their own physical fog computing servers (e.g. the high-end industrial routers), neither require them to upgrade their IoT equipment to compliant to fog computing; the management team can invoke the public fog to their system in order to distribute the tasks from their central server to the public fog in the proximity of their front-end IoT devices. Certainly, such an environment simplifies the establishment of fog computing and ideally reduces the cost from installing and maintaining physical equipment. However, it also raises a new challenge regarding the work assignment optimisation. Here, Figure~\ref{fig:scen} illustrates a Mobile Big Data acquisition \cite{Chang2018:MobileBigData} scenario that expresses the challenge.

In this scenario, a Big Data Service Provision (BDSP) system allows its client to request on-demand sensory data collection and pre-processing based on the customised algorithm of the client. Suppose the Big Data Service Provision Server (BDSP Server) has received a request from a Bid Data Service client (BDS Client) who intends to collect semi-real time information within the same district based on processing the collected sensory data using the client's algorithm written in the script supported by the BDSP. After BDSP Server validates the request package, it creates and dispatches the Process Deployment Package (PDP) to a Data Source Server (DSS) hosted on a mobile sensor located in the same district as the BDS Client. The request involves collecting 100 sensory data objects in a period of time and process the sensory data objects one by one using the client's algorithm. Although the distant compute servers (co-located with the BDSP Server) are always available for reducing the resource usage of DSS in performing the data processing, in order to reduce the latency, BDSP Server has commanded DSS to migrate the process to proximal fog servers managed by their partner companies whenever they are available. Initially, when DSS received the PDP from BDSP Server, it has discovered Fog1 and Fog2 servers. However, when the sensory data is ready to be processed, DSS has moved to another location where Fog1 and Fog2 are no longer connectable, and the new discovered servers are Fog3, Fog4 and Mist1 \cite{Liyanage2018IJPCC}. Since the fog servers may be serving the other clients, their resource states are dynamically changing all the time. 

The scenario raises two questions:

\begin{itemize}
    \item \textit{What is the optimal approach to migrate DDS's tasks to the heterogeneous fog servers?}
    \item \textit{How to realise such dynamic process deployment and execution on DSS?}
\end{itemize}

In order to answer the questions, we propose Resource-aware Edge process Migration (REM) scheme which is capable of assigning works based on runtime hardware states of the participants. Further, in order to realise such a system and to evaluate the proposed scheme, we developed Edge Process-enabled Internet of Things (EPIoT) host, which is an evolved software framework from the classic embedded Web server \cite{Srirama2006:MWS}. Further, we have implemented the proposed solution and have tested it on real-world equipment. In general, the results have shown that the proposed scheme is capable of improving the overall speed of the process distribution in the heterogeneous fog and edge computing environment.

This paper is organised as follows. Section 2 summarises the related works and specifies the differences between the proposed work and the past related works. Follow up in Section 3, the authors describe the proposed EPIoT framework and the involved components. Afterwards, in Section 4, the authors explain the proposed REM scheme and provide the details of the prototype implementation and the evaluation of the REM scheme in Section 5. Finally, this article is concluded in Section 6 together with future research directions.

%% file: sections/relatedWorks.tex
\section{Related Works}

The massive increase of connected ubiquitous and smart devices in our lives have been leading in the development of data streaming and cloud computing applications, such as computer vision~\cite{Zubal2016,Lopez2018}, augmented reality \cite{Alam2017}, security \cite{Levi2017}, real-time monitoring \cite{Kresimir2016} and tracking \cite{Luca2015}. Consequently, the Internet of things (IoT) constitutes a major fabricator of big data which leads to two major issues for computing IoT-generated data. First one is related to the processing time since there are so many factors that can affect the response time such as the network delay and the performance of the available processing power. The second issue is related to the data uploading from a large number of IoT devices which can introduce network congestion and network delay. Therefore, it is important to know how to partition the data among all the available resources in the network for better performance in the processing time. Accordingly, many researchers have engaged in investigating the data partitioning and stream processing for analytics purposes. For example, Yang \cite{Yang2017} has proposed a general model and architecture of fog data streaming based on analyzing common properties of numerous applications. Specifically, the approach took into consideration four important dimensions: system, data, human, and optimization. Correspondingly, the results were promoting the combination of network edge and stream processing. 

Another simple and widely spread approach is focusing on duplicating the data. Specifically, the concept is literally based on replicating the data in order to strengthen and reinforce the key characteristic of an efficient system: its availability, its increasing fault-tolerance, its low bandwidth consumption, and its enhanced scalability. Thus, many works addressed this point and proposed methods with low-cost and high data availability protocol. For example, Deris et al. \cite{deris2008efficient} presented a box-shaped grid structure that helps to maintain the consistency of the replicated data on a networked distributed computing systems.  Further, the outcome of their approach demonstrated a high data availability with a low communication cost which influenced positively the fault-tolerance with respect to baseline protocols. 

In the same category, pipsCloud~\cite{wang2018pipscloud} tries to address the increasing requirement for reliable and up-to-date information about the resources and the environment. Specifically, the proposed method focuses on using cloud computing model with High-Performance Computing (HPC) techniques to obtain large-scale multi-spectral remote sensing datasets and processing. The approach was designed for optimal querying and fast accessing. The results have shown that the algorithm is capable of performing efficiently. 

With a focus on the network capacities, there is always this issue of network skew that can cause workers to struggle in completing the tasks. As a solution, Rupprecht et al. \cite{Lukas2017} presented a \textit{SquirreJoin} approach based on distributed join processing technique using \textit{lazy partitioning} in which the algorithm helps in adapting to transient network skew in clusters. In detail, the process starts by maintaining lazy partitions in memory by the workers, then these partitions are assigned dynamically to other workers based on the network conditions. In general, the whole principle allows this ability to distribute the data flow and to minimize the join completion time. Particularly, the solution was capable of speeding up 2.9x times performance with only small overhead. 

Yang et al. \cite{Lei2013} have also introduced a solution for optimizing the data partitioning where the focus was on computation partitioning, which means that the optimization is done on the partitioning of the data stream. Specifically, the authors proposed a framework to provide runtime support for a dynamic partitioning of the data and sharing the computational instances among multiple users. In addition, the framework is characterized by its flexibility to serve a large number of mobile users by leveraging the existing elastic resources. Particularly, the partitioning problem was solved by using a genetic algorithm which allowed the framework to have 2X improvement on the performance.

The optimization of data partitioning can present some additional challenges when the data itself is with higher dimensions which are the case in IoT applications \cite{Marjani2017}. This means that the data has to be fused in order to ensure a good quality of the information. Therefore, the researchers focused on data fusion techniques as a mean to fuse the data and facilitate the process of improving efficiency and computational performance. This aspect is clearly stated and illustrated in the work of Zhou et al. \cite{Zhou2013}, where the authors proposed a multidimensional fusion approach for IoT data based on partitioning. Specifically, the concept relies on splitting the size of the data with high attributes into small subsets for processing and then apply reduction and extraction methods to fuse the results together. Further, a simulation was executed for testing the correctness of the algorithm which has proven to be effective.

With this respect, the rise of IoT offers this possibility of producing a massive amount of data with possibilities of processing in it at different locations and levels which makes data partitioning strategy equivalent to NP problem. Hence,  Naas et al. \cite{Naas2018} tried to propose a solution to the issue based on dividing the original data problem into subproblems using graph modelling and partitioning methods. The resulting heuristics makes it possible to reduce the solving time by 450 times with 5\% optimality loss. 

All the methods and approaches discussed in this section show that most of the methods proposed for partitioning and processing are relying on the databases management for data partitioning with  Cloud resources or IoT within their horizontal networks. Further, although a number related works \cite{Marinelli2009:Hyrax,Fernando2012:Honeybee,chang2014spica,Loke2015,Soo2017} in Mobile Cloud Computing (MCC) \cite{fernando2013mobile} domain have proposed solutions for distributing data processing tasks among devices, they have either not considered the heterogeneity of the participative computational resources or have not included resources across different layers from the cloud to the edge. Therefore, in this paper, we are presenting a new design for partitioning and processing the data by taking into account all available resource in the vertical and horizontal layers of the network.

%% file: sections/system.tex
\section{System Design}

This section describes the proposed system for dynamic process deployment and execution. Follow up, we explain the proposed REM scheme used for performing optimal process migration among heterogeneous servers.

\subsection{System Architecture}
In this paper, we call the activities performed to fulfil the required tasks at the front-end IoT devices as Edge Network Entities-Assigned Processes or simply Edge Processes. Here, the edge represents the network where the front end IoT devices located and near edge denotes the intermediate network between the network of the front-end and the distant backend central server.

\begin{figure}[h]
  \centering
    \includegraphics[width=0.98\textwidth]{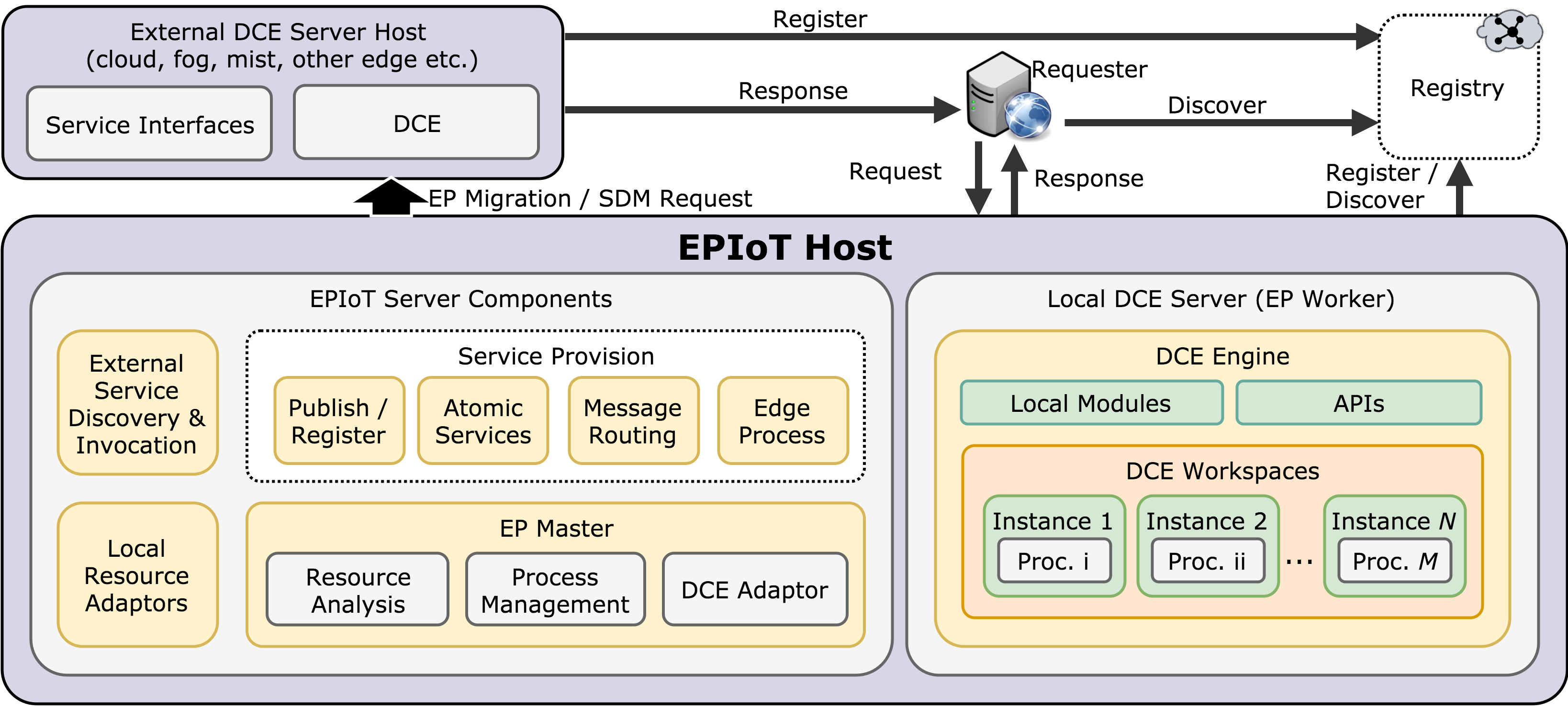}
    \caption{Software architecture of EPIoT host. An EPIoT host contains two main elements - EPIoT Server, which is the manager of EPIoT host, and DCE Server, which handles EP work packages.}
    \label{fig:EPIoT}
\end{figure}

In this system, dynamic process deployment is the core enabler, it is realised by implementing a corresponding engine for Dynamic Code Execution (DCE).

DCE is a mechanism that allows a client to send the package, which may contain the configuration description file, program source code files and optional dependencies, module or library files, to the DCE server and the DCE server is capable of executing the program on-the-fly using real-time code compiler. Note that the DCE client may require the DCE server to download the dependencies instead of sending them to the server directly.

Commonly, a system can provide DCE by utilising virtualisation or containerisation technologies. Similarly, we expect the participative servers of the dynamic fog computing environment (dynamic fog) are providing DCE mechanism.
In a dynamic fog-enabled IoT system, we call a DCE-supported IoT device as Edge Process-enabled Internet of Thing (EPIoT) host.
EPIoT host is discoverable via both registry servers and proximity-based wireless communication protocols (e.g. Wi-Fi or Bluetooth). It is accessible to an organisation and its partners as long as the corresponding software agent has been installed in the participants. It may have different hardware states at different period of time depending on the missions they are carried. Hence, it periodically updates its Service Description Metadata (SDM) to the registry via the Internet.

An EPIoT host contains two main parts---EPIoT Server and DCE Server.

EPIoT Server contains the basic mechanisms of a service-oriented IoT device together with additional components that allow it to manage DCE tasks. EPIoT Server has following service provision related components:
\begin{itemize}
    \item \textbf{Publish/Register} component monitors hardware states and manages service availability. It needs to include hardware states in the Service Description Metadata (SDM) of EPIoT Host when it publishes it to the registry. Further, Publish/Register component is capable of identifying the availability of the services based on the corresponding hardware states. For example, if EPIoT Server has received a request that requires continuous network transmission for a period of time, it may stop providing the Message Routing Service in order achieve better quality for the current task.
    \item \textbf{Atomic Services} component provides simple services such as sensory data service, which allows the requester to use EPIoT host to collect real-time sensory data, or actuating service that allows the requester to access the actuators connected with the EPIoT host. For example, a connected mobile actuator, which embeds EPIoT, allows distant control centre to use it to access a wireless light switch in the proximity of the mobile actuator.
    \item \textbf{Message Routing} service component is a service that supports the basic function of Software-Defined Networking (SDN). Specifically, an IoT system can utilise multiple EPIoT host devices to establish a dynamic re-configurable sensory data routing network. Further, internal components can also use Message Routing to route messages to external nodes. For example, if the description metadata of an edge process service request package has specified the output receiver to be a different node from the initial requester, EP Master will use Message Routing to route the output to the receiver node.
    \item \textbf{Edge Process} service component allows the requester to send a request package to EPIoT host and execute the program to fulfil the request in a secure sandbox environment. In particular, the request package should contain the source code of the program, required dependencies/library for the program (in case they are not available in the EPIoT host) and the metadata which specifies the process configuration details. For example, the metadata should describe which file contains the starting point of the program, which sensory data is needed (in case the program involves pre-processing the data collected by the EPIoT host), the receiver of the final output (in case if the requester and receiver are different nodes) etc.
\end{itemize}

\textbf{External Service Discovery \& Invocation} component allows EPIoT host:
\begin{itemize}
    \item To discover external servers via global federated registry and it also allows the EPIoT host to discover proximity-based servers using proximity-based wireless service discovery mechanisms such as Wi-Fi, Wi-Fi Direct, Bluetooth etc. We assume the available external servers are following an industrial standard format to describe their services and their resource availability. For example, they may follow ETSI Multi-access Edge Computing API standard\footnote{see \url{https://www.etsi.org/technologies-clusters/technologies/multi-access-edge-computing}} and discoverable using IETF RFC 6762 - Multicast DNS  with IETF RFC 6763 - DNS Based Service Discovery.
    \item To send the process migration package to external servers such as the cloud, the fog, edge computing or mist computing servers. Similarly, the request package should include the program, dependencies, the configuration metadata and the involved data objects (if the request involves them).
    \item To request Service Description Metadata (SDM) from external DCE server and also managing the SDM of EPIoT host itself.
\end{itemize}

\textbf{Local Resource Adaptors} are the adaptors for other entities to access the basic functions of EPIoT host device (which fundamentally is an IoT device) which are accessing sensory data and interacting with actuators.

\textbf{EP Master} handles the request package of edge process service. Specifically, it contains the following main mechanisms:
\begin{itemize}
    \item \textit{Resource Analysis} function allows EP Master access the information of the underline hardware and network resource. For example, when EP Master intends to migrate the processes it received from the requester, it uses Resource Analysis function to identify the current CPU, RAM and networking states towards performing adaptive process migration mechanism.
    \item \textit{Process Management} function allows EP Master:
    \begin{itemize}
        \item To extract the content of an edge process request package and adding the required data (if the request involves sensory data collected by the EPIoT host) to create a work package. Afterwards, Work Manager sends the work package to the DCE server for execution.
        \item To partition the work package and to create a number of new edge process packages when EPIoT host intends to distribute the work to external servers. Further, as mentioned previously, Work Manager can send the edge process packages to external servers via the External Service Invocation component. EP Manager partitions the work based on DIPHDA scheme, we will describe it in the next section.
    \end{itemize}
    \item \textit{DCE Adaptor} is a component that allows EP Master to deploy work packages to DCE Engine. A different system may use different type DCE engine, in order to deploy process packages to the DCE engine, a corresponding adaptor needs to be installed in EP Master so that the format of process/work package is compliant to the DCE engine.
    
\end{itemize}

\textbf{Local Dynamic Code Execution (DCE) Server}. In general, every node in the distributed fog computing has hosted a DCE engine and the engine is managed by an isolated server. The EPIoT host device can install DCE server to allow itself processing the work locally. This server is an independent server from the EPIoT Server because in different cases, the administrator may prefer to use a different type of DCE, such as Virtual Machine (VM) or containers engine (e.g. Docker). Hence, this part does not bound with EPIoT Server.

\textbf{DCE Engine} is a component that is capable of receiving process package and executing the process. Afterwards, the DCE engine will return the output to the process package sender.

Due to the security reason, DCE engine executes each process package in an isolated sandbox environment. Further, the executed program of the process package has limited accessibility to Local Resource Adaptors via the pre-installed Application Program Interface (API) of the engine. Certainly, the corresponding information about API is also included in the SDM of EPIoT host.

DCE engine may also provide a number of Local Modules (i.e. pre-installed dependencies or libraries) for the programming language it supports. Fundamentally, edge process service requester cannot expect EPIoT host to support all the dependencies they need. Hence, the edge process request packages should include the corresponding dependencies/libraries for their programs. Otherwise, the EPIoT host may need to spend extra time to download the dependencies/libraries from external sources.

\subsection{Process Migration}
The EPIoT host is capable of migrating the work it received to external DCE-enabled servers. In this case, the role of EPIoT host becomes Master Node or Delegator Node and the roles of external DCE servers will be Worker Nodes.

Below, we use Business Process Management Notation (BPMN) model to explain how a Master Node migrates its works to a Worker Nodes in which the Worker Node is a Fog Computing server hosted on the Wi-Fi access point connected by the Master Node.

\begin{figure}
  \centering
    \includegraphics[width=1.0\textwidth]{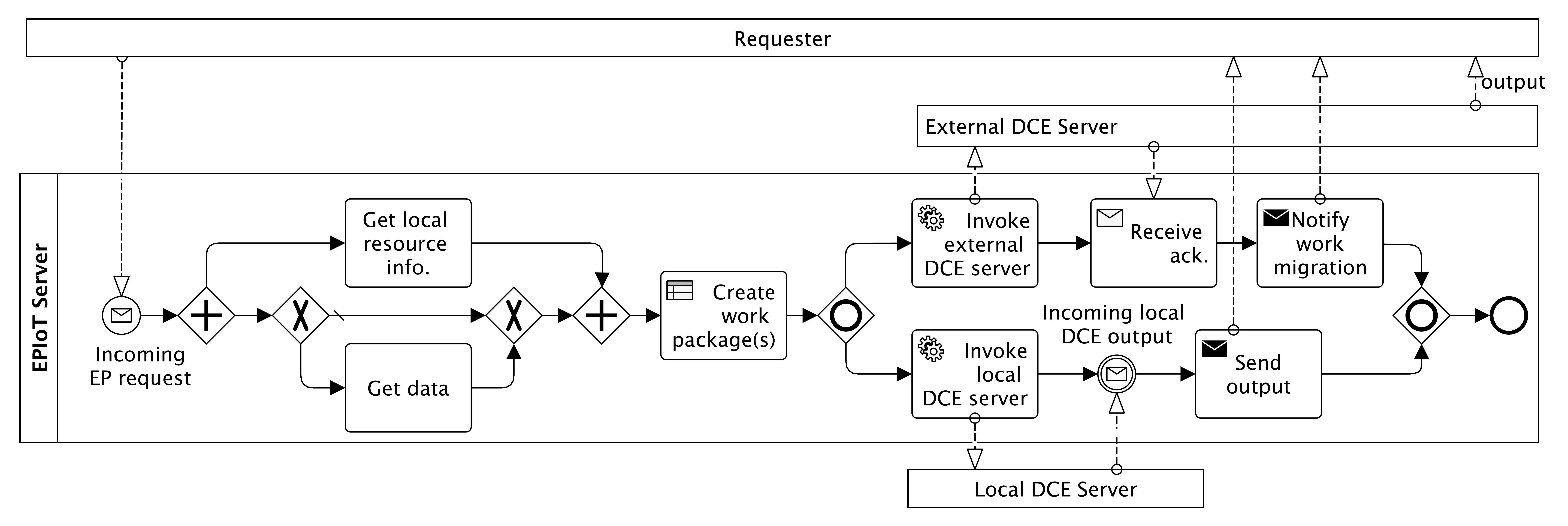}
    \caption{The workflow of handling edge process service request. When EPIoT host receives Edge Process (EP) service request, it may handle the request using local DCE server or it may migrate the work to external DCE servers.}
    \label{fig:Migration}
\end{figure}

As Figure~\ref{fig:Migration} shows, when the EPIoT host receives EP request, it first checks the local resource states and in parallel, if the request involves sensory data, it will get the data either by utilising adaptors to collect on-demand data or get the data from local storage, depending on the requested content. Afterwards, the EPIoT host will create process packages. Each process package contains the original files included in the request package, and the data retrieved in the previous step (if involved). Further, EPIoT host allocates the data to each process. For example, suppose the request involves 100 sequel sensory data and every 10 sequel data is the input of one process, which indicates that there will be 10 processes in total.  Based on the rules set by the system admin or the manager of the EPIoT host device, there can be three cases:

\begin{enumerate}
    \item EPIoT host creates one process package and sends it to local DCE server. Afterwards, receive the output from local DCE server and forward the output to the requester.
    \item EPIoT host creates one or more process package and invokes external DCE servers to send them the process packages. When an external DCE server receives a completed process package, it will send an acknowledgement to EPIoT host. Afterwards, EPIoT host notifies the initial requester that the process has been migrated to external servers and EPIoT host also provide the information (e.g. IP address, identification etc.) about the external servers. This case is known as process migration. Based on the metadata included in the work package, the external DCE servers know the address of the initial requester. Hence, they can directly send the final outputs to the initial requester.
    \item EPIoT host creates multiple process packages and sends the process packages to both local DCE server and external DCE server(s). In this hybrid case, the system intends to use all the possible resources to hasten the overall speed and in addition, reduce the burden of the EPIoT host.
\end{enumerate}

The decision of how EPIoT assigns the process packages is based on the proposed resource-aware edge process migration scheme explained in the next section.

%% file: sections/rem_scheme.tex
\section{Resource-aware Edge Process Migration Scheme}
The proposed Resource-aware Edge process Migration (REM) scheme is capable of identifying the optimal partition for migrating processes from EPIoT node to heterogeneous external servers situated at the cloud and the edge network resources.

REM scheme requires four sets of parameters in order to measure the optimal partition:
\begin{enumerate}
    \item Static resource specification of the participants. Here, the participants include the EPIoT host in which the EPIoT server of the EPIoT host is the master node or the delegator node exchangeably, and the local DCE server of EPIoT host and the external DCE servers are worker nodes. To enumerate, the information of the static resource specification include the CPU benchmark value, RAM size, disk read and write speed.
    \item Dynamic context parameters, which include the current network transmission speed between EPIoT host and the participants, current CPU usage, current RAM usage, which are the parameters that can influence the overall speed of the entire operation. Here, we consider the entire edge process migration operation encompassed deployment, which includes packing the process package and sending the package to the internal or external DCE server, process execution timespan and the timespan of sending the output to the receiver.
    \item Request parameters, which include the following values:
        \begin{itemize}
            \item $byte_{alg}$---bytes of the algorithm source code, which is visible when EP Master receives the EP request package.
            \item $byte_{mdl}$---bytes of modules, which is visible when EP Master receives the EP request package.
            \item $byte_{desc}$---bytes of the description metadata, which describes the program configuration and the address of receiver node and it is visible when EP Master receives the EP request package.
            \item $byte_{d}$---bytes of one data object involved in the request. EPIoT knows the size of the data since it is from one of the localhost component.
        \end{itemize}
    \item Local process time. In order to perform the measurement, EPIoT host has to perform at least one process once locally. Specifically, one process involves: (1) packing the process package, which include the algorithm source code, dependancies, metadata and one of the requested data; (2) sending the process package to local DCE server for execution and generating output package; (3) sending the output package to the receiver. Afterwards, EPIoT host can obtain the values described in Table~\ref{table:notations}.
\end{enumerate}

\begin{table}[]
\centering
\caption{Description of the notations involved in measuring local process time.}
\label{table:notations}
\scriptsize
\begin{tabular}{|l|p{10cm}|}
\hline
Notation & Description                                                                                                                                             \\[5pt] \hline\hline
$\Delta t^{upk_{mdl}}_{local} $        & Timespan to unpack modules/dependancies.                                                                                                                \\[5pt] \hline
$\Delta t^{upk_{alg}}_{local}$        & Timespan to unpack Algorithm/main program.                                                                                                              \\[5pt] \hline
$\Delta t^{pk_{mdl}}$        & Timespan to pack modules/dependancies.                                                                                                                  \\[5pt] \hline
$\Delta t^{pk_{alg}}$        & Timespan to pack Algorithm/main program.                                                                                                                \\[5pt] \hline
$\Delta t^{pk_{d}}$        & Timespan to pack 1 data object.                                                                                                                         \\[5pt] \hline
$\Delta t^{upk_d}_{local}$        & Timespan to unpack 1 data object.                                                                                                                       \\[5pt] \hline
$\Delta t^{o1}_{local}$        & Timespan to pack 1 output data.                                                                                                                         \\[5pt] \hline
$\Delta t^{proc1}_{local}$        & Timespan to process 1 data object.                                                                                                                      \\[5pt] \hline
$\Delta t^{npk_{o1}}_{local}$        & Timespan to unpack 1 archived output. Note that this is an additional trial that EPIoT needs to perform in order to measure the timespan of the output. \\ \hline
\end{tabular}

\end{table}

REM scheme performs in the steps described below.
Firstly, REM scheme scores each server based on the values described previously by using the function $getTime(i, wp_i)$ where $i$ denotes worker $i$ and $wp_i$ denotes the number of data object assigned to worker $i$.

Let $WT = \{ wt_i : 0 \leq i \leq N \}$ be a set of the total timespan of each worker $i$ after assigning $wp_i$ to it.

REM scheme uses the process assignment function (i.e. Algorithm~\ref{func:assignment}) to assign works to the workers.

\RestyleAlgo{boxruled}
\LinesNumbered
\begin{algorithm}[H]
\caption{Process assignment function}
\label{func:assignment}
\SetAlgoLined
\For{$i \in WP$}{
  $wp_i = 0$
}
\While{$\# D > 0$}{
  \For{$wt_i \in WT$}{
    \If{$wt_i \equiv min\{wt_i\}$}{
      $\#D = \#D - 1$\;
      $wp_i = wo_i + 1$\;
      $wt_i = getTime(i, wp_i)$\;
    }
  }
}

\end{algorithm}

In Algorithm~\ref{func:assignment}:
\begin{itemize}
\item $min\{wt_i\}$ is a function that returns the smallest number in $WT$.
\item $getTime(i,wp_i)$ is a function in which:
\begin{equation}
\begin{aligned}
    getTime(i,wp_i) =  &\Delta t^{pk}_{i} + \Delta t^{req}_{i} + \Delta t^{upk}_i  
   + \Delta t^{proc}_i \\ &+ \Delta t^{pk_O}_i + \Delta t^{po_O}_i + \Delta t^{npk_O}_r
\end{aligned}
\end{equation}
\end{itemize}

where:
\begin{itemize}
    \item $\Delta t^{pk}_{i}$ is the time to pack the request package assigned to worker $i$ in which:
    	\begin{equation}
        	\Delta t^{pk}_{i} = \Delta t^{pk_{mdl}} + \Delta t^{pk_{alg}} + \Delta t^{pk_{d}} \times wp_i
        \end{equation}
        where:
        \begin{itemize}
        \item $\Delta t^{pk_{d}} \times wp_i$ is the data object packing time based on the number of assigned data objects.
        \end{itemize}
    %
    %
    \item $\Delta t^{req}_{i}$ is the time to send the request package to worker $i$ in which:
    \begin{equation}
    	\Delta t^{req}_{i} = \Delta t^{po_{1byte}}_i \times (byte_{mdl} + byte_{alg} + byte_{d} \times wp_i )
    \end{equation}
    where:
    \begin{itemize}
    \item $byte_{d} \times wp_i$ is the size of data object in byte times number of data object assigned to the worker.
    \end{itemize}
    %
    %
    \item $\Delta t^{upk}_{local}$ is the timespan when the request package is handled by the delegator itself in which:
    \begin{equation}
    	\Delta t^{upk}_{local} = \Delta t^{upk_{mdl}}_{local} + \Delta t^{upk_{alg}}_{local} + \Delta t^{upk_d}_{local} \times wp_i
    \end{equation}
    %
    %
    \item $\Delta t^{upk}_i$ is the measured process package unpacking time of worker $i$, which is measured based on Read/Write speed in which:
    \begin{equation}
    	\Delta t^{upk}_i = \frac{1}{\Delta t^{upk}_{local}} \times \frac{RW_i}{RW_{local}}
    \end{equation}
    where:
    \begin{itemize}
    \item $RW_{local}$ is the average read and write performance of the delegator in which:
    	\begin{equation}
        	RW_{local} = \frac{R_{local} + W_{local}}{2}
        \end{equation}
    \item $RW_i$ is the average read and write performance of worker $i$ in which:
    	\begin{equation}
        	RW_i = \frac{R_i + R_w}{2}
        \end{equation}
    \end{itemize}
    %
    %
    \item $\Delta t^{proc}_{local}$ is the timespan when the the delegator executes the process by itself in which:
    	\begin{equation}
        	\Delta t^{proc}_{local} = \Delta t^{proc1}_{local} \times wp_i
    	\end{equation}
    \item $\Delta t^{proc}_i$ is the measured timespan for worker $i$ to execute the process in which: 
    \begin{equation*}
        \Delta t^{proc}_i = 
        wp_i
        \times
        \frac{1}{\Delta t^{proc}_{local}}
        \times
        \frac{
            \frac{1}{\sum_{k\in |H_{i}|} \omega_k} \sum_{k\in |H_{i}|} \eta_k^{i} \times \omega_k
        }{
            \frac{1}{\sum_{k\in |H_{local}|} \omega_k} \sum_{k \in |H_{local}|} \eta_k^{local} \times \omega_k
        }
    \end{equation*} and it can also be written as:
    \begin{equation}
        \Delta t^{proc}_i = 
        \frac{wp_i}{\Delta t^{proc}_{local}}
        \times
        \frac{\sum_{k\in |H_{i}|} \eta_k^{i} \times \omega_k}{\sum_{k\in |H_{i}|} \omega_k}
        \times
        \frac{\sum_{k\in |H_{local}|} \omega_k}{\sum_{k \in |H_{local}|} \eta_k^{local} \times \omega_k}
    \end{equation} where:
    \begin{itemize}
        \item $\omega_k$ is the weight of the computational resource---$k$. Initially, the weights of all the resources are equal (e.g. weight = 1). The system administrator can configure a threshold that increases the weight value.
        \item $\eta_k^{local}$ is the value of the computational resource---$k$ of delegator.
        \item $\eta_k^{i}$ is the value of the computational resource---$k$ of worker $i$.
    \end{itemize}
    %
    %
    \item $\Delta t^{pk_O}_{local}$ is the output packing time when the delegator is handling it, in which:
    	\begin{equation}
        	\Delta t^{pk_O}_{local} = \Delta t^{o1}_{local} \times wp_i
        \end{equation}
    %
    %
    \item $\Delta t^{pk_O}_i$ is the output packing time when worker $i$ is handling it. Specifically, it is based on the disk read and write speed, in which:
    	\begin{equation}
        	\Delta t^{pk_O}_i = \frac{1}{\Delta t^{pk_O}_{local}} \times \frac{RW_i}{RW_{local}}
        \end{equation}
    \item $\Delta t^{po_O}_i$ is the measured timespan to send result to the output receiver described in the metadata of the request package. Further, this value is based on the network speed of the worker $i$ (described in its Service Description Metadata) and the network speed of receiver (provided in the metadata in the request package) and also the size of the output package.
    \item $\Delta t^{npk_O}_r$ is the timespan to unpack the output package at the receiver side and  it is measured as below: \begin{equation}
        \Delta t^{upk_O}_r = 
        \frac{1}{\Delta t^{upk_O}_{local}} \times
        \frac{RW_i}{RW_{local}}
    \end{equation}
\end{itemize}

%% file: sections/evaluation.tex
\section{Evaluation}
In order to evaluate the proposed EPIoT framework and the proposed REM scheme, we have implemented a prototype software and have deployed it on real world devices. First, we explain the configuration of the test cases which is based on the scenario described in Figure 1 of Section 1.

\begin{itemize}
\item Data Source Server (DSS), which is the EPIoT host, is simulated by a Nokia 8 smartphone. It contains all the components of EPIoT host described in Figure~\ref{fig:EPIoT}.
\item Mist server has been simulated by a Sony Xperia XZ1 Compact, which is discoverable and communicable via Wi-Fi Direct.
\item The two Fog nodes (i.e. Fog 3 and Fog 4) in Figure~\ref{fig:EPIoT} have been simulated by a Macbook Pro model A1502 (Fog 3) and a Dell Latitude E5430 laptop (Fog 4). Specifically, Fog 3 is the Internet gateway of DSS via IEEE802.11n.
\item Additionally, we have included a geo-distributed cloud in our testing. The cloud  is a virtual machine located in the datacenter of Tallinn University of Technology.
\end{itemize}

\begin{table}[h]
\caption{Static specifications and configurations of the participative nodes in evaluation.}
\label{table:setting}
\centering
\scriptsize
\begin{tabular}{|p{2cm}|p{1.4cm}|p{1.4cm}|p{1.4cm}|p{1.4cm}|p{1.4cm}|}
\hline
& DSS/EPIoT host                                     & Mist                                     & Fog 3
& Fog 4
& Geo-distributed Cloud                                     \\ \hline\hline
Simulated by           
& Nokia 8 smartphone                         & Sony Xperia XZ1 Compact smartphone         & Macbook Pro A1502     
& Dell Latitude E5430           
& VM                                         \\ \hline
OS                       
& Android                                     & Android                                     & Mac OS                
& Ubuntu                        
& Debian                                     \\ \hline
CPU
& Qualcomm MSM8998 Snapdragon 835
Octa-core (4x2.5 GHz Kryo \& 4x1.8 GHz Kryo)
& Qualcomm MSM8998 Snapdragon 835
Octa-core (4x2.45 GHz Kryo \& 4x1.9 GHz Kryo)
& Intel Core i5 2.6 GHz 
& Intel Core i5-3210M; 4x2.5GHz 
& QEMU Virtual CPU version 2.5+;
4x2.40GHz 
\\ \hline
Available CPU core 
& 2                                         & 2                                         & 3                     
& 3
& 3                                           \\ \hline
RAM                      
& 4 GB                                       & 4 GB                                       & 16 GB                 
& 4 GB                          
& 8 GB                                       \\ \hline
Disk speed               
& 547 MB/s (R); 220 MB/s (W)                 & 237 MB/s (R);121 MB/s (W)                   & 566 MB/s (R/W)        
& 105 MB/s (R/W)                
& 61 MB/s (R/W)                               \\ \hline
Communication            
& Localhost                                   & Wi-Fi Direct                               & Wi-Fi 802.11n         
& via Node-F                    
& via Node-F                                 \\ \hline
\end{tabular}
\end{table}


Table~\ref{table:setting} summarises the specification of the nodes.

Note that although the CPU specifications of the nodes are 4 or more cores, we have configured a limited core to the DCE servers hosted on those nodes towards simulating the constraint resource availabilities of the nodes.  

Besides the static values, participative nodes have to collect the dynamic values such as CPU usage, RAM usage, network speed at runtime. In general, each node is collecting such information at the background and periodically update such information to their Service Description Metadata (SDM). Further, EPIoT host retrieves SDM from the other nodes when it connects to the network.

Regarding the software prototype, we have implemented EPIoT host and DCE server using Node.js. Further, we installed the Node.js-based EPIoT host on Nokia 8 via Termux (Linux command prompt emulator for Android OS)\footnote{see \url{https://termux.com/}} and all the nodes are using the same DCE server software. Similarly, we also installed DCE server on Sony Xperia XZ1 Compact via Termux.

The sensory data objects used in test cases are image files of different sizes and the algorithm program source code included in the request package is 1203 bytes. Further, the size of the modules/libraries used by the algorithm program is 14.1 megabytes and the size of the configuration metadata (in JSON format) is 226 bytes.

The following evaluation consists of three parts - the first part is the preliminary testing that focuses on the plain testing of local process and mono-process migration, which shows the performance when EPIoT handles the processes locally or when it migrates its process to one single external server; the second part of the evaluation aims to demonstrate how much the proposed REM scheme can optimise the process migration when EPIoT host considers to migrate its processes to multiple heterogeneous nodes; the final part tends to show that the proposed REM scheme can improve the speed in all the possible multi-process migration cases.

\subsection{Preliminary Testing}
This section describes the test cases for local processes and mono-process migrations. In particular, the involved sensory data object size was 1 megabyte and we have used a different number of data objects in different test cases.

Note that the following test results have excluded the timespan of the initial request. Specifically, the figures have only contained the timespan of Deploy, which consists of packing the process migration package, sending the package to the worker node and the unpacking time of the worker node, and the Process \& Response represents the program execution time on the worker node, output packing time and the time to send the output to the receiver node, which in our test cases is the initial requester.

\subsubsection{Local Process}
\begin{figure}[h]
  \centering
    \includegraphics[width=0.50\textwidth]{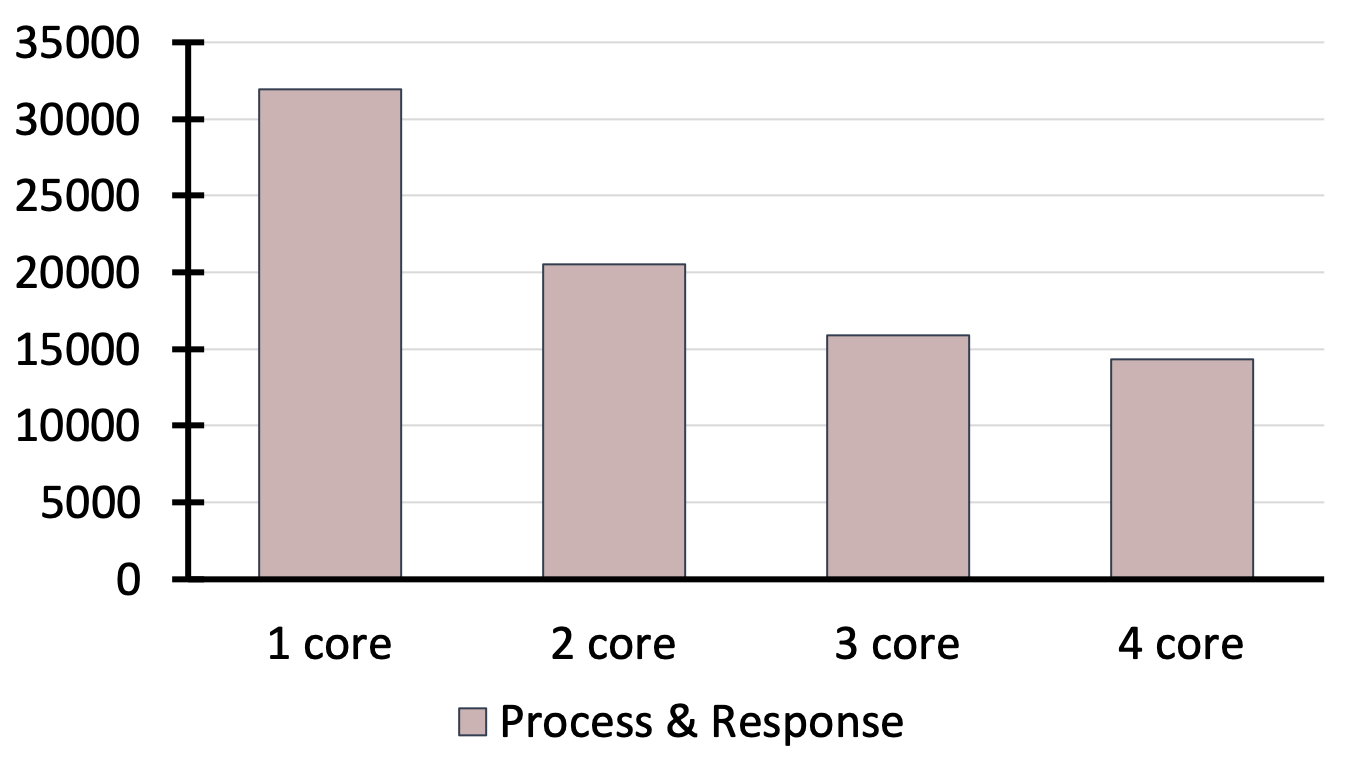}
    \caption{Edge process performance on local server based on the number of allocated CPU resources with 50 of 3 megabyte data object included in the deployment package.}
    \label{fig:localProcess}
\end{figure}

Figure~\ref{fig:localProcess} illustrates the test cases for local processes performed on EPIoT host based on 50 involved data objects per one request package and the number of available CPU core assigned to the local DCE server of the EPIoT host.

The main objective of these test cases is to show that EPIoT host device itself is only capable of allocating a limited number of CPU cores to the local DCE server. Explicitly, when we force the system to allocate more than 3 CPU cores to the local DCE server, DCE server could not reduce the timespan for executing the processes significantly.

In the rest of the test cases, we only allocate 2 CPU core to the DCE server of EPIoT host and the CPU core allocations of the other nodes have followed the setting in Table~\ref{table:setting} described previously.

\subsubsection{Mono-Process Migration}

\begin{figure}[h]
  \centering
    \includegraphics[width=0.98\textwidth]{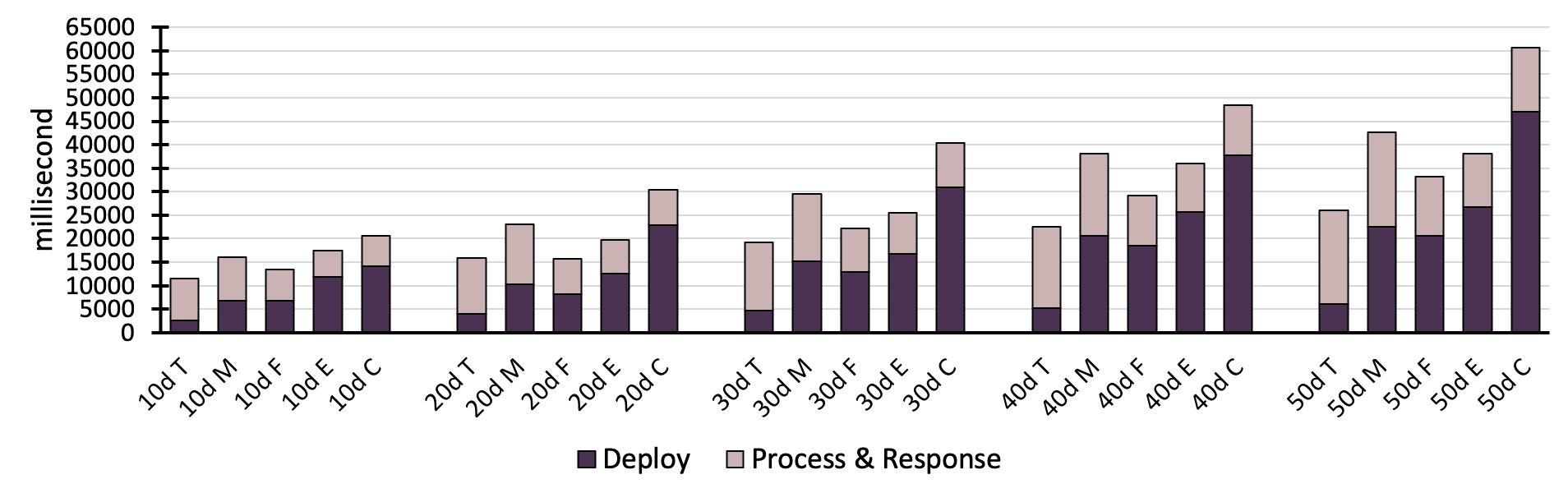}
    \caption{Time comparison for local processes and mono-process migration with different number of 3 megabytes data objects included in the process deployment package.
}
    \label{fig:monoMigration}
\end{figure}

Figure~\ref{fig:monoMigration} illustrates the results of the test cases for mono-process migration based on a different number of data objects included in the initial request. To enumerate, 10d denotes 10 data objects, 20d denotes 20 data objects and so forth. Further, we use a single alphabet, which corresponding to the nodes described previously, to represent the node that was handling the processes. Specifically, T denotes DSS/EPIoT node; M denotes Mist server node; F denotes the Fog 3 node; E denotes the Fog 4 node and  C denotes the geo-distributed cloud node.

As the figure shows, EPIoT host (T) has the lowest deployment time because EPIoT server only needs to send the process package to the local DCE server. Follow up by Fog 3 (F), which is the main Internet gateway of EPIoT host and is accessible via IEEE 802.11n. In comparison, Mist (M) has higher latency than Fog 3 because the performance of Wi-Fi Direct is slower than the regular Wi-Fi IEEE 802.11n connection in these test cases. Further, Fog 4 (E) has higher latency than Fog 3 since it is the 2nd hop from EPIoT host and finally, the geo-distributed cloud (C), which is the cloud server located in another city, has the highest latency among all the nodes.

Regarding the computational performance, Fog 3 (F), Fog 4 (E) and geo-distributed cloud (C) have quite similar results and EPIoT host and Mist (M) have very similar performance since both of them have similar CPU chip.

\subsection{REM Scheme-based Process Migration}

This section describes the test cases that validate the efficiency of the proposed REM scheme. Specifically, the test cases encompass two parts in which the first part of test cases validate the REM scheme using the requests that involve different number of same size data objects; the second part of test cases validate the REM scheme using different size of data objects with a fixed number of data objects involved in the request.

\subsubsection{Performance Influenced by Number of Data Objects}

The test cases of this part aims to show that how much the REM scheme can improve the overall performance in comparison with local processes and the na\"ive process migration approach which equally divide the data objects to the process migration packages regardless the runtime context (e.g. CPU load, RAM usage, network speed etc.) of the worker nodes.

\begin{figure}[h]
  \centering
    \includegraphics[width=0.98\textwidth]{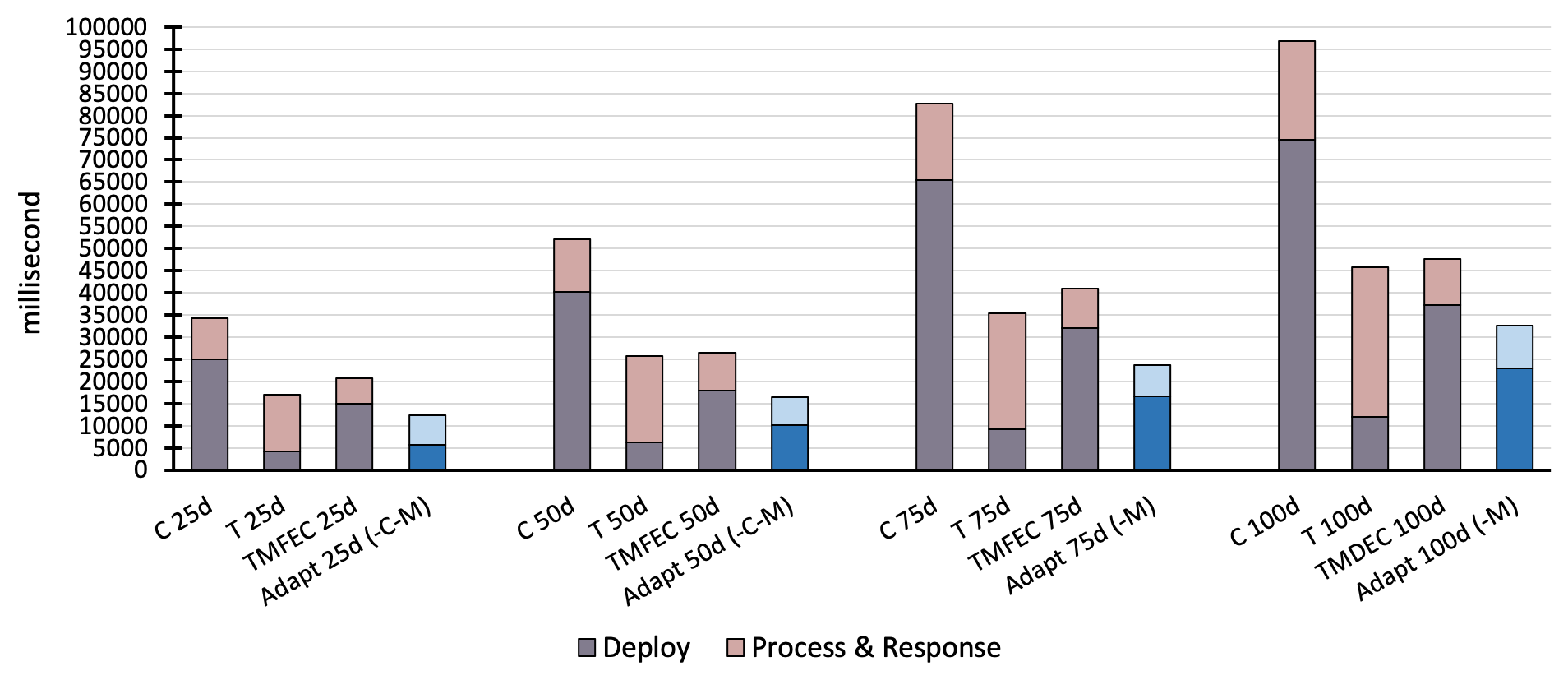}
    \caption{Deployment, process and response timespan comparison among solo, equal and REM-based process migration using multiple worker nodes. Individual data object size is 3 megabyte.}
    \label{fig:25d100d}
\end{figure}

Figure~\ref{fig:25d100d} illustrates the results of the test cases that involved 25 to 100 data objects (denoted by 25d, 50d, 75d, 100d) in the request. Specifically, each set of test cases (i.e. based on the number of the involved data objects) consists of one case of mono-process migration to the geo-distributed cloud (denoted by C), one case of local process (denoted by T), one na\"ive case which equally migrate the processes to all nodes including localhost (denoted by TMFEC) and the last case of a set represents the result that have applied REM scheme (denoted by Adapt).

As the results have shown, na\"ive approaches can improve the performance comparing to the mono-process migration to the cloud. However, the case that is based on the local process can outperform the na\"ive approach. In contrast, when the system utilised REM scheme, the system has adaptively assigned the data objects to the processor package and has also excluded the nodes that cloud reduce the performance from the candidate workers. Hence, in all the sets of test cases, REM scheme-based approach has outperformed all the other approaches. Further, we have included the annotations in order to clarify which worker nodes have been excluded by the REM scheme. For example, -C denotes REM scheme has excluded the cloud node and -M denotes REM scheme has excluded the Mist node.

\subsubsection{Performance Influenced by The Size of Data Objects}

In contrast to the previous part of test cases, the test cases described in this section were based on the size of the data objects while the request package has involved 25 data objects.

\begin{figure}[h]
  \centering
    \includegraphics[width=0.98\textwidth]{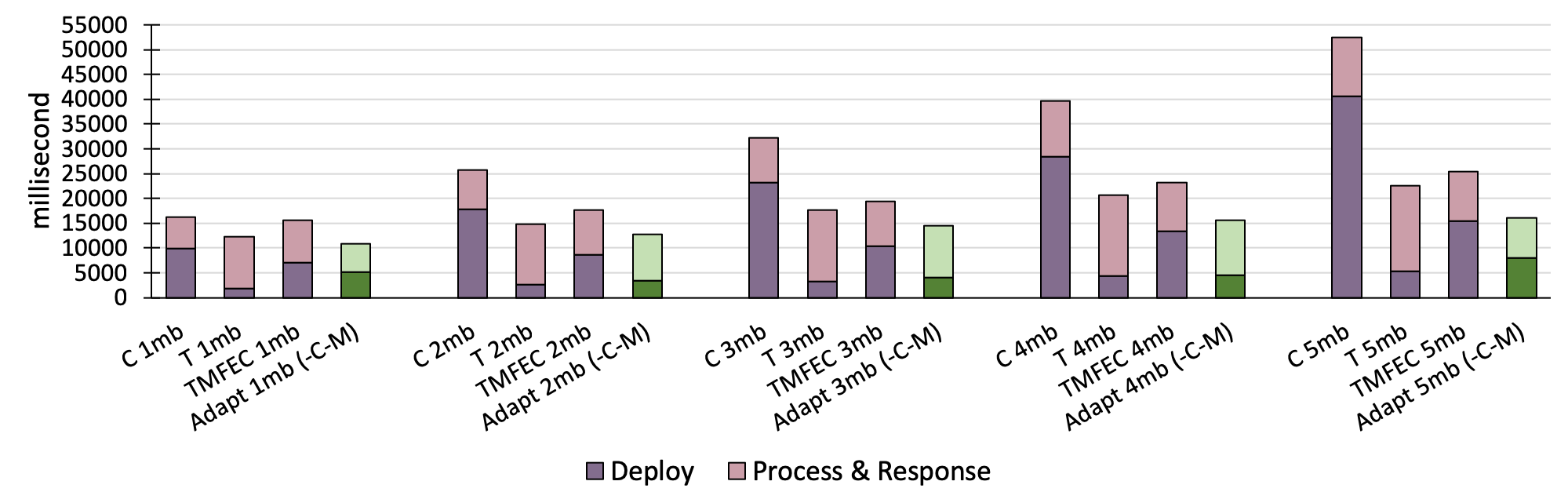}
    \caption{Deployment, process and response timespan comparison among cloud-based process migration, solo processing, equal process migration and adaptive process migration. The number of requested data object is 25.}
    \label{fig:25xNmb}
\end{figure}

Similarly to the results of the previous section, the cloud-based mono-process migration has suffered from the network latency even though its processing performance was better than the EPIoT host device. Moreover, the na\"ive approach (denoted by TMFEC) also did not perform well and the local process-based approach (denoted by T) has outperformed the na\"ive approach. Explicitly, the REM scheme-based approaches have outperformed the other approaches in all the sets of test cases.

Note that REM scheme improves the performance based on all the runtime context factors. Hence, it does not always reduce one part of timespan. For example, when the data object size was 4 megabyte per each, REM scheme-based approach has reduced the network latency in order to improve the performance. In contrast, when the data object size was 5 megabyte, REM scheme-based approach produced a different option in which it results that the network latency was much higher in comparison to the case that involved 4-megabyte data. However, REM scheme's decision has reduced the timespan of processing. Hence, the overall performance has still been improved.

\subsection{Applying REM Scheme in Different Computing Models}

In fog and edge computing paradigms, there are many different possible approaches to selecting the nodes for process migration or process distribution. In order to further validate the REM scheme, we have conducted all the possible setting in our testing environment. Correspondingly, we use a single alphabet to represent a node in which T denotes EPIoT node, M denotes Mist node, F denotes Fog 3 node, E denotes Fog 4 node and C denotes the geo-distributed cloud node. Further, we use "A." to notate the cases that have applied REM scheme.

\begin{figure}[h]
  \centering
    \includegraphics[width=0.99\textwidth]{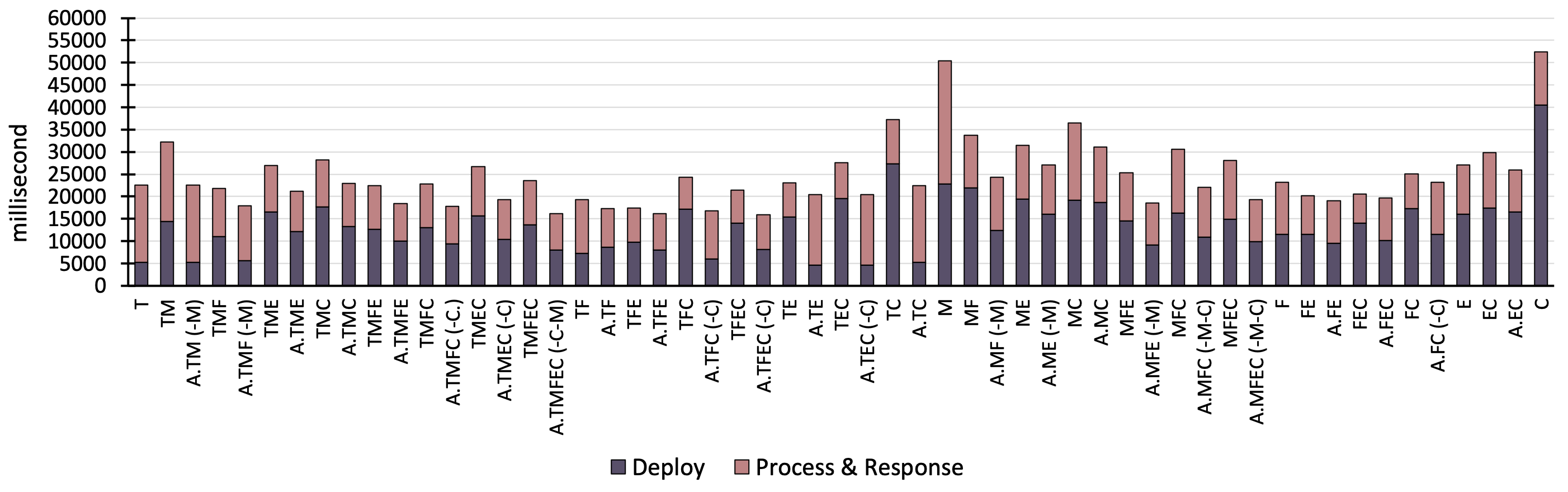}
    \caption{Deployment, process and response timespan comparison among different cases. REM scheme is capable of improving the performance of all the cases. The number of requested data object is 25 with the size of 5 megabyte per each data object.}
    \label{fig:compareAll}
\end{figure}

Figure~\ref{fig:compareAll} illustrate the results of the test cases that have 25 involved data objects in the size of 5 megabytes per each data object. Note that we have performed the test cases of each section in a different period of time. Hence, due to the context changes, the timespans of test cases can be different in different sections of this paper.

In these test cases, we can consider T as edge computing or Things Computing, M and as mist computing, C as cloud computing, F as fog computing, E as fog or edge computing and the rest test cases are the combination of different elements. For examples, TM, which utilises EPIoT host and Mist node, would be a common paradigm of mist computing and TFE, which utilises EPIoT host (T), Fog 3 (F) and Fog 4 (E) would be a common paradigm of fog computing and so forth.

As the results have shown, REM scheme can improve the performance of all the test cases that involve more than one participative process executioner. Similar to the previous sections, we have annotated the test cases in which REM scheme has excluded a number of candidate nodes.

%% file: sections/conclusion.tex
\section{Conclusion and Future Work}
In this paper, the authors have proposed a scheme and a framework to address two questions in the domain of the distributed process migration from mobile IoT devices (connected mobile sensors) to heterogeneous Fog and Edge Computing (FEC) resources. Specifically, in order to improve the performance of the process migration, the authors proposed Resource-aware Edge process Migration (REM) scheme. Further, the authors have developed and have implemented Edge Process-enabled Internet of Things (EPIoT) host framework, which has been utilised to evaluate the REM scheme. Overall, based on the experimental testing on real-world equipment and devices, the authors have shown that the proposed REM scheme is capable of improving the performance of different types of FEC settings that contains heterogeneous resources and dynamic runtime context factors.
For the future work, the authors plan to extend the work with the following mechanisms.
\begin{itemize}
\item Enabling generic Software-Defined Networking (SDN) service from the EPIoT host. Current EPIoT host supports message routing mechanism. However, it is not yet fully compliant with the mechanisms of SDN in which the nodes should be able to manage routing tables and is capable of reducing the traffic by eliminating redundant bytes of routing packages.
\item Integrates the proposed EPIoT framework and REM scheme with the Intelligent Transport System (ITS) \cite{Plangi2018} in order to enhance the speed of the real-time process management and decision making. Further, the authors plan to implement and to validate the framework based on real-world use cases.
\end{itemize}